\documentclass[a4paper,fleqn,usenatbib,useAMS]{mnras}

\usepackage{newtxtext,newtxmath}

\usepackage[T1]{fontenc}
\usepackage{ae,aecompl}

\usepackage{graphicx}	
\usepackage{amsmath}	
\usepackage{multicol}   
\usepackage{amssymb}	
\usepackage{booktabs}   

\title[Binarity of hybrid pulsator KIC~5709664]{Spectroscopic confirmation of the binary nature of the hybrid pulsator KIC~5709664 found with the frequency modulation method}

\author[A. Derekas et al.]{
A.\ Derekas,$^{1,2,3}$\thanks{E-mail: derekas@gothard.hu}
S.\ J.\ Murphy,$^{4,5}$
G.\ D\'alya,$^{6}$
R.\ Szab\'o,$^{2,17}$
T.\ Borkovits,$^{7,2}$
A.\ B\'okon,$^{8}$
\newauthor
H.\ Lehmann,$^{9}$
K.\ Kinemuchi,$^{10}$
J.\ Southworth,$^{11}$
S.\ Bloemen,$^{12}$
B.\ Cs\'ak,$^{2,1}$
H.\ Isaacson$^{13}$
\newauthor
J.\ Kov\'acs,$^{1,3}$
A.\ Shporer,$^{14}$
Gy.\ M.\ Szab\'o,$^{1,3}$
A.\ O.\ Thygesen,$^{15}$
Sz.\ M\'esz\'aros$^{1,3,16}$
\\
$^{1}$ ELTE E\"otv\"os Lor\'and University, Gothard Astrophysical Observatory, Szombathely, Hungary\\
$^{2}$ Konkoly Observatory, Research Centre for Astronomy and Earth Sciences, Hungarian Academy of Sciences, H-1121 Budapest, \\
Konkoly Thege Mikl\'os \'ut 15-17, Hungary\\
$^{3}$ MTA-ELTE Exoplanet Research Group, 9700 Szombathely, Szent Imre h. u. 112, Hungary\\
$^{4}$ Sydney Institute for Astronomy, School of Physics, The University of Sydney, NSW 2006, Australia\\
$^{5}$ Stellar Astrophysics Centre, Department of Physics and Astronomy, Aarhus University, Ny Munkegade 120, DK-8000 Aarhus C, Denmark \\
$^{6}$ Institute of Physics, E\"otv\"os University, 1117 Budapest, Hungary\\
$^{7}$ Baja Astronomical Observatory of Szeged University, H-6500 Baja, Szegedi \'ut, Kt. 766, Hungary \\
$^{8}$ Department of Experimental Physics, University of Szeged, H-6720 Szeged, D\'om t\'er 9, Hungary\\
$^{9}$ Th\"uringer Landessternwarte Tautenburg, Karl-Schwarzschild-Observatorium, 07778 Tautenburg, Germany\\
$^{10}$ Apache Point Observatory, Sunspot NM 88349, USA\\
$^{11}$ Astrophysics Group, Keele University Newcastle-under-Lyme, ST5 5BG, UK\\
$^{12}$ Department of Astrophysics/IMAPP, Radboud University Nijmegen, P.O. Box 9010, NL-6500 GL Nijmegen, Netherlands\\
$^{13}$ Department of Astronomy, University of California at Berkeley, Berkeley, CA 94720-3411, USA\\
$^{14}$ Kavli Institute for Astrophysics and Space Research, M.I.T., Cambridge, MA 02139, USA\\
$^{15}$ California Institute of Technology, 1200 E. California Boulevard, Pasadena, CA 91125, USA\\
$^{16}$ Premium Postdoctoral Fellow of the Hungarian Academy of Sciences\\
$^{17}$ MTA CSFK Lend\"ulet Near-Field Cosmology Research Group\\
 \\
}

\date{Accepted XXX. Received YYY; in original form ZZZ}

\pubyear{2018}

\begin{document}
\label{firstpage}
\pagerange{\pageref{firstpage}--\pageref{lastpage}}
\maketitle

\begin{abstract}
We started a program to search for companions around hybrid $\delta$\,Sct/$\gamma$\,Dor stars with the frequency modulation method using {\it Kepler} data. Our best candidate was KIC~5709664, where we could identify Fourier peaks with sidelobes, suggesting binary orbital motion. We determined the orbital parameters with the phase modulation method and took spectroscopic measurements to confirm unambiguously the binary nature with radial velocities. The spectra show that the object is a double-lined spectroscopic binary, and we determined the orbital solutions from the radial velocity curve fit. The parameters from the phase modulation method and the radial velocity fits are in good agreement. We combined a radial velocity and phase modulation approach to determine the orbital parameters as accurately as possible. We determined that the pulsator is a hybrid $\delta$\,Sct/$\gamma$\,Dor star in an eccentric binary system with an orbital period of $\sim$95\,d and an eccentricity of 0.55. The measured mass ratio is 0.67.  We analysed the pulsation content and extracted 38 frequencies with amplitudes greater than 20\,$\mu$mag. At low frequencies, we found broad power excesses which are likely attributed to spots on the rotating surface of the lower-mass component. We inferred rotation periods of 0.56 and 2.53\,d for the primary and secondary, respectively.
\end{abstract}




\section{Introduction} \label{sec:intro}

Thanks to very precise observations from the recent generation of space telescopes (such as CoRoT and {\it Kepler}), we know that almost all stars show some type of pulsation. It is also known that more than 50\% of stars are in binary or multiple systems \citep{alf14}. Since binarity can influence the pulsation properties in different ways, it is very important to identify companions around pulsating stars if we aim for the determination of very accurate stellar properties.

Until recently, there were two photometric methods to discover companions of pulsating stars: observation of eclipses, or the O$-$C method in the case of mono-periodic pulsating stars. Of course, radial velocities (RVs) give unambiguous evidence for binarity. \cite{shibahashi&kurtz2012} developed a method to find companions of multi-periodic pulsating stars in frequency space, which they called the frequency modulation (FM) method, and which makes the detection and study of these systems easier. In the case of a binary system in which at least one of the objects is a variable star with coherent oscillations, all of the relevant parameters of the system can be calculated using the frequency spectrum of the light curve: the orbital period, the semi-major axis and even the RVs and the mass function which are traditionally extracted from spectroscopic observations. The star and its companion orbit around the common centre of mass, hence the distance that the light from the star has to travel to the telescope is periodically shorter and longer than in the case where there is no companion. This phenomenon is called the light-time-effect, and manifests itself as a frequency modulation in the Fourier spectrum of the light curve of the star, causing multiplets around every frequency peak.

Besides the FM method, \citet{murphy&shibahashi2015} and \citet{murphyetal2016b} expanded and developed the phase modulation (PM) method \citep{murphyetal2014}. Phase and frequency modulation are equivalent, and so the PM method can also be used for detecting binary systems using the stellar oscillations and calculating their orbital parameters. Because the PM method involves binning of the light curve, it is sensitive only to systems where the period is greater than twice the length of bins.

Following the ground-breaking work of \cite{shibahashi&kurtz2012}, we started an investigation of hybrid $\delta$\,Sct/$\gamma$\,Dor stars classified by \cite{uyt11} using long cadence data of the {\it Kepler} space telescope to find companions around them. These are A/F type main-sequence stars with coherent oscillations, lying in the instability strip of the Hertzsprung-Russell Diagram. They simultaneously show two types of non-radial pulsations: $\delta$\,Sct low-order p\:modes with periods 0.008$-$0.42\,d, and $\gamma$\,Dor low-degree high-order g\:modes with periods 0.3$-$3\,d \citep{lam18,san18}. One of the biggest spectroscopic surveys of these hybrid pulsators was performed by \cite{lam18}, who investigated the binary fraction of 50 such hybrid pulsators based on RV measurements. They detected companions for 27\% of their sample. \cite{mur18} investigated 2224 main sequence A/F stars and found 341 non-eclipsing binaries with the PM method.

We studied 585 $\delta$\,Sct, $\gamma$\,Dor and hybrid $\delta$\,Sct/$\gamma$\,Dor stars in our binary search project with the FM method. When it started in 2012, only {\it Kepler} Q0-Q10 data were available. One of the best candidates was KIC~5709664, which showed a consistent side-peak pattern at pulsation frequencies with the largest amplitudes. We then started to collect spectroscopic data to confirm its binarity.

The brightness of KIC~5709664 is Kp=11.2\,mag. Its basic physical parameters are T$_{\rm eff}$=6820\,K, $\log g$=4.4\,dex, [Fe/H]=-0.2\,dex and v$\sin i$= 71.4\,km\,s$^{-1}$ based on the spectra from the APOGEE Data Release 14 \citep{abo18}. \cite{mur18} identified it as a non-eclipsing binary using the PM method and determined the orbital parameters from the pulsation ($P_{\rm orb} = 95.03$\,d; $a_{1}\sin{i}=50.5$\,R$_{\odot}$; $e=0.51$; $K = 23.1$\,km\,s$^{-1}$) and listed it as a single-pulsator binary.

Here, we present the photometric and spectroscopic analysis of KIC~5709664. In Sect.\ \ref{sec:observ}, we describe the {\it Kepler} data used in the photometric analysis and the spectroscopic observations we took. In Sect.\ \ref{sec:results}, we present our results of the photometric and spectroscopic orbit determination. The frequency analysis is  discussed in Sect.\ \ref{sec:freqs} and finally we summarise our results in Sect.\ \ref{sec:sum}.

\section{Observations} \label{sec:observ}

We used the {\it Kepler} data in the photometric analysis of KIC~5709664. A detailed description of the telescope and data processing can be found in \cite{bor10}, \cite{gil10}, \cite{jen10a,jen10b} and \cite{koc10}. KIC~5709664 was observed in long-cadence mode (30-min exposures) for the full 4-yr mission (Quarters 0--17), while short-cadence observations (60-s exposures) were taken in Q2.2 and the whole of Q5. In this paper, we use all quarters (Q0--17) of the long-cadence observations, only. These constitute 65313 data points that have a time span of 1470.5\,d and a duty cycle of 90.8\%.

\begin{figure*}
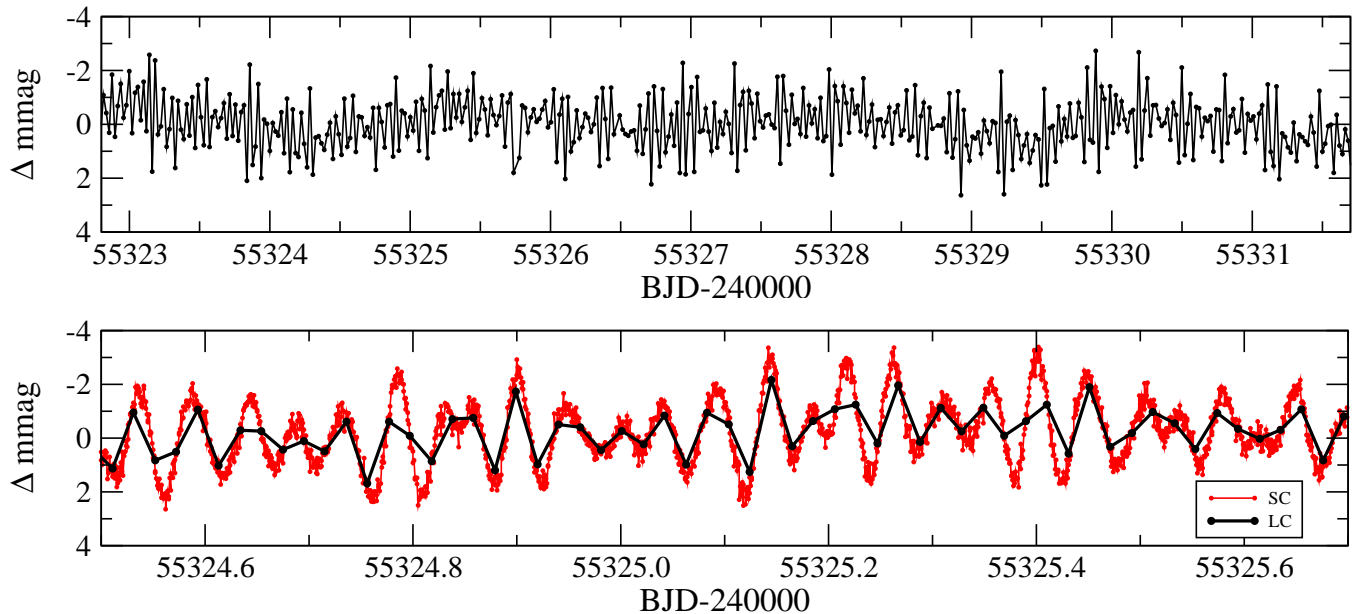

\includegraphics[width=\linewidth]{lc3s.eps}\\[5pt]
\includegraphics[width=\linewidth]{lc1day.eps}
\caption{\label{lc} Top panel: a 9-d segment of the light curve. Bottom panel: 1-d segment of the light curve. The red symbols and line show the short cadence (60s), the black symbols and line the long-cadence data (30 min).}
\end{figure*}

We used the multi-scale MAP light curve \citep{stumpeetal2014}, obtained from KASOC.\footnote{\url{http://kasoc.phys.au.dk}}. A 9-day section of the light curve is plotted in Fig.\ \ref{lc}.

We obtained medium- and high-resolution spectra at six observatories. The average S/N of these spectra was 40-80. The observations are described below and summarised in Table\:\ref{obsjour}.
\begin{itemize}
\item We took one spectrum in 2012\,May with the eShel spectrograph mounted on a 0.5-m Ritchey-Chr\'etien telescope at the Gothard Astrophysical Observatory (GAO), Szombathely, Hungary, in the spectral range 4200--8700\,\AA\ with a resolution of R = 11\,000.
\item We obtained three spectra in 2012 June with the 4.2~m William Herschel Telescope at La Palma using the ISIS spectrograph. It has dichroic filters which permit simultaneous observations in the blue and red arms. The blue arm used the H2400B grating and covered the 4170--4570\,\AA\ wavelength range at a resolution of 12\,000. The red arm used the R1200R grating and covered 6040--6830\,\AA\ at a resolution of 8500.
\item We obtained two spectra at Apache Point Observatory (APO), USA, using the ARCES \'echelle spectrograph on the 3.5~m telescope with a resolution of R = 31\,500 in the spectral range 3200--10\,000\,\AA. One spectrum was taken in 2012 October, and another in 2018 April.
\item We took nine spectra with the 2\,m Alfred Jensch Telescope at Th\"uringer Landessternwarte Tautenburg in August and September 2016. The spectrograph was used with a projected slit width of 2 arcsec providing a resolving power of R = 30\,000.
\item We used spectra taken by the Apache Point Observatory Galactic Evolution Experiment (APOGEE; \citealt{maj17}), part of the 3rd and 4th Sloan Digital Sky Survey \citep{eis11,bla17}. APOGEE is a high resolution (R=22,500) near-infrared survey, that is observing nearly 500,000 stars between 2011 and 2020 in the wavelength range of 15090-16990\,\AA\ \citep{wil12}. We used 5 spectra from the 14th data release of SDSS \citep{abo18} which were taken in 2013 September and October.
\item  We collected five spectra using the KECK I telescope equipped with the HIRES instrument with resolution of 60\,000. The data were collected without the iodine cell. Spectra were taken between 2016 June and November and covered the 3640--7960\,\AA\ wavelength range. The technical information of the HIRES setup is described in \citep{shporeretal2016}.
\end{itemize}

\begin{figure*}
\includegraphics[width=\linewidth]{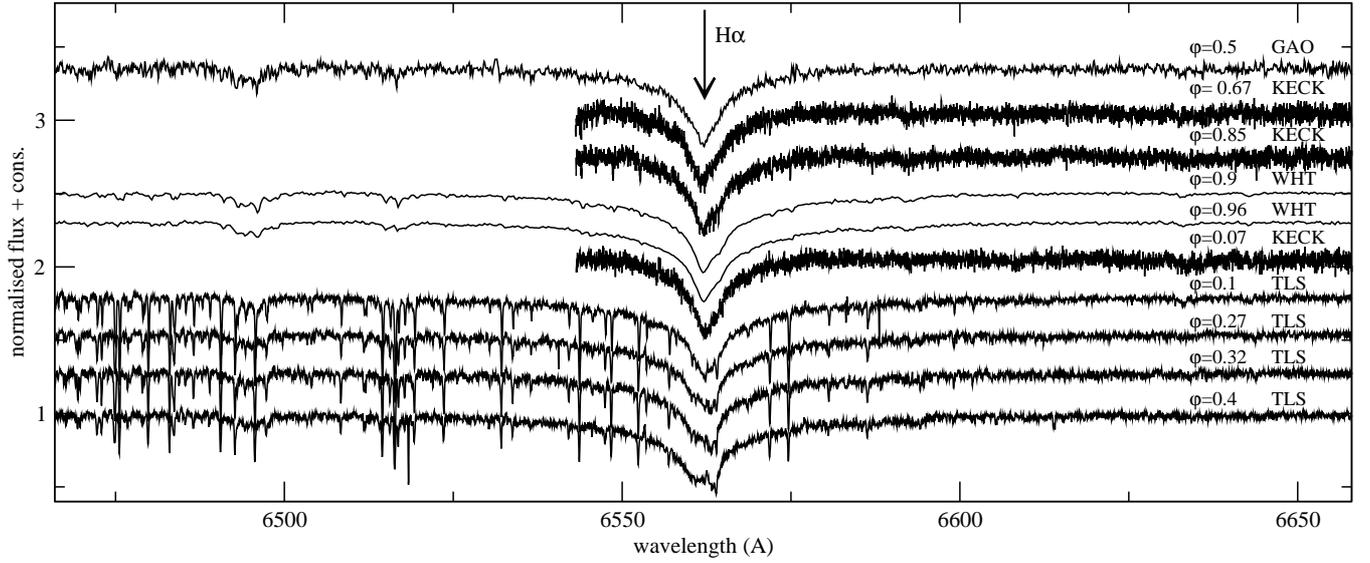}
\caption{Selected spectra at different orbital phases, showing the effect of binary motion in the H$\alpha$ line. The phases (as inferred in Sect.~\ref{sec:results}) and the observatories are given above each spectrum.}
\label{fig:spectra}
\end{figure*}

All spectra were reduced either using {\sc iraf} or a dedicated pipeline, then normalized to the continuum level. Except data from the APOGEE survey, the radial velocities were determined by cross-correlating the spectra with a well-matched synthetic template spectrum from the extensive spectral library of \citet{mun05}.  We used the 3850--6850\,\AA\ region in the cross-correlation, except for a 20\,\AA\ region around the Sodium D lines. Since the stellar rotation is fast, all the lines are broadened, so the spectra are dominated by the hydrogen lines. In Fig.\,\ref{fig:spectra}, we show spectra focusing on the H$\alpha$ line, illustrating the effect of the binary motion at different orbital phases.

All RVs were corrected to barycentric RVs. The APOGEE data are reduced by a dedicated pipeline, {\sc aspcap} \citep{gar16}, which derives atmospheric parameters, rotation velocities, and chemical abundances by comparing observations with libraries of theoretical spectra. The measured RVs from all spectroscopic sources are listed in Table\ \ref{radveldata}.

\begin{table}
\caption{Journal of observations.}
\label{obsjour}
\begin{tabular}{llcc}
\hline
Observatory & \multicolumn{1}{l}{Wavelength range} & Resolution  & No. of  \\
            & \AA\        &       &  spectra\\
\hline
WHT & 4170--4570 and 6040--6830  & 12\,000 and 8500 & 3  \\
GAO & 4200--8700 & 11\,000 & 1 \\
APOGEE & 15\,090--16\,990 & 22\,500   & 5\\
APO & 3200--10\,000 & 31\,500 & 2 \\
TLS & 4540--7540 & 30\,000     & 9  \\
KECK &  3640--7960  &  60\,000    & 5 \\

\hline
\end{tabular}
\end{table}

\begin{table}
\begin{center}
\caption{Radial velocity measurements. HJD is shown without the prepending `24'.}
\label{radveldata}
\begin{tabular}{lrrc}
\toprule
\multicolumn{1}{c}{HJD} &  \multicolumn{1}{c}{RV$_{1}$} &  \multicolumn{1}{c}{RV$_{2}$} & Observatory \\
 & \multicolumn{1}{c}{km\,s$^{-1}$} &  \multicolumn{1}{c}{km\,s$^{-1}$} &  \\
\midrule
56048.53828 & $-20.4 \pm 0.9$ & $-12.8 \pm 1.6$ & GAO \\
56086.73189 & $-44.6 \pm 1.8 $ & $-16.1 \pm 3.0$ & WHT \\
56089.55007 & $-39.1 \pm 1.7 $ & $-9.0 \pm 2.8$  & WHT \\
56092.55340 & $-38.6 \pm 1.8$ & $-16.5 \pm 2.8$ & WHT \\
56228.66885 & $44.6 \pm 0.6$ & \multicolumn{1}{c}{---} & APO \\
56557.73272 & $-40.4 \pm 0.8$ & \multicolumn{1}{c}{---} & APOGEE \\
56559.72264 & $-37.8 \pm 0.6$ & \multicolumn{1}{c}{---} & APOGEE \\
56560.72038 & $-38.5 \pm 0.5$ & \multicolumn{1}{c}{---} & APOGEE \\
56584.63151 & $-22.3 \pm 0.6$ & \multicolumn{1}{c}{---} & APOGEE \\
56585.63001 & $-20.8 \pm 0.5$ & \multicolumn{1}{c}{---} & APOGEE \\
57569.08281 & $-24.2 \pm 0.5$ & $-19.9 \pm 0.5 $ &     KECK \\
57583.93289 & $-44.2 \pm 0.5$ & $-15.4 \pm 0.5 $ &   KECK \\
57601.08965 & $-41.7 \pm 0.5$ & $-10.3 \pm 0.5 $ &   KECK \\
57621.94122 & $-29.8 \pm 0.5$ & $-18.7 \pm 0.5 $ &   KECK \\
57622.34135 & $-31.8 \pm 0.5$ & $-12.4 \pm 1.0$ & TLS \\
57624.42425 & $-30.3 \pm 0.6$ & $-15.2 \pm 1.0$ & TLS \\
57624.45236 & $-30.6 \pm 0.5$ & $-15.2 \pm 0.9$ & TLS \\
57641.38582 & $4.7 \pm 0.5$ & $-38.6 \pm 1.0$ & TLS \\
57642.37919 & $8.4 \pm 0.5$ & $-41.4 \pm 1.0$ & TLS \\
57643.47454 & $12.5 \pm 0.5$ & $-45.1 \pm 1.4$ & TLS \\
57644.48464 & $15.7 \pm 0.6$  & $-48.6 \pm 1.5$ & TLS \\
57645.49054 & $15.7 \pm 0.5$ & $-51.9 \pm 1.3$ & TLS \\
57653.34960 & $44.9 \pm 0.4$ & $-63.8 \pm 1.3$ & TLS \\
57718.72473  &  $-28.9 \pm 0.6$ & $-18.6 \pm 0.6$ & KECK \\
58214.98802 & $17.6 \pm 0.6$ & $-54.5 \pm 1.1$ & APO \\
\bottomrule
\end{tabular}
\end{center}
\end{table}

\section{Results}\label{sec:results}

\subsection{Photometric orbit determination}

\begin{figure}
\includegraphics[width=\columnwidth]{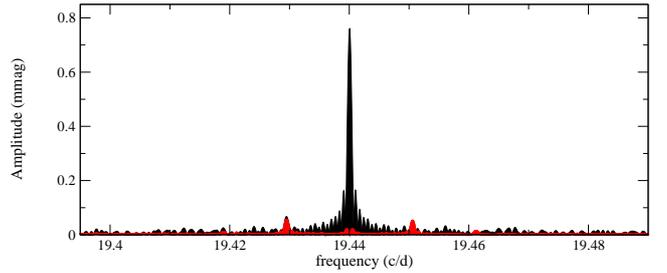}
\caption{Fourier transform of the light curve at the strongest oscillation frequency (black). Strong sidelobes exist as a result of binary motion, separated from the main peak by the orbital frequency. At twice the orbital frequency from the main peak, smaller sidelobes are evident, indicating an eccentric orbit \citep{shibahashietal2015}. The orbital sidelobes remain even after prehitening the pulsation peak (red).}
\label{fig:peak}
\end{figure}

 The binary orbit induces frequency modulation on the pulsation frequencies \citep{shibahashi&kurtz2012}, generating sidelobes on each peak in the Fourier transform (Fig.\,\ref{fig:peak}). In order to obtain a set of excited, observed oscillation frequencies without sidelobes, we had to correct the observation times to the barycentre of the binary system. This first requires a photometric measurement of the orbital parameters, for which we used the PM method \citep{murphyetal2014,murphy&shibahashi2015}.

\begin{figure}
\includegraphics[width=\columnwidth]{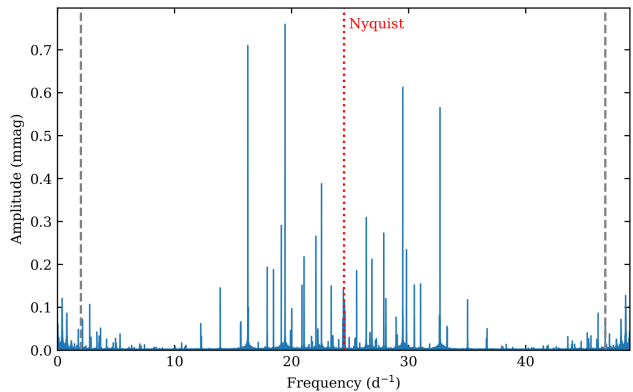}
\caption{The Fourier spectrum of KIC~5709664 based on the 4-yr LC {\it Kepler} dataset. The dotted red line is the Nyquist frequency. Some real peaks lie above the Nyquist frequency, and can be distinguished from their aliases by amplitude. The dashed grey lines delimit the frequency region considered for frequency extraction (see Sec.\,\ref{sec:freqs}).}
\label{fig:fourier}
\end{figure}

The Fourier transform of the {\it Kepler} light curve of KIC~5709664 is dominated by p\:modes near the long-cadence Nyquist frequency of 24.48\,d$^{-1}$ (Fig.\,\ref{fig:fourier}). There is also variability at low frequencies, whose spectral window contributes noise at all frequencies and degrades the quality of a PM orbital solution. We therefore followed the methodology of \citet{murphyetal2016a} and high-pass filtered the light curve. It is important to note that we did this only for the PM analysis, and subsequent analysis uses an unfiltered light curve, as described later in this section.

\begin{figure*}
\includegraphics[width=\linewidth]{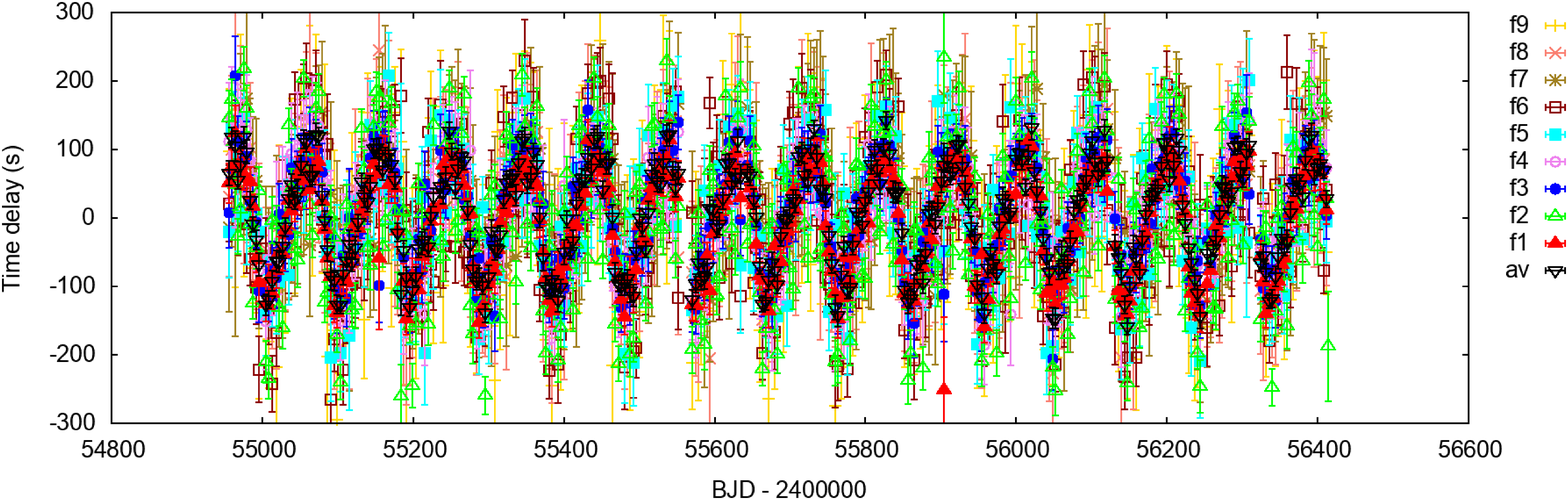}
\caption{Time delays for the nine strongest modes of KIC~5709664 and the weighted average time delay (black), at 4-day sampling in the full 4-yr light curve.}
\label{fig:td_err_bar}
\end{figure*}

\begin{figure}
\includegraphics[width=\columnwidth]{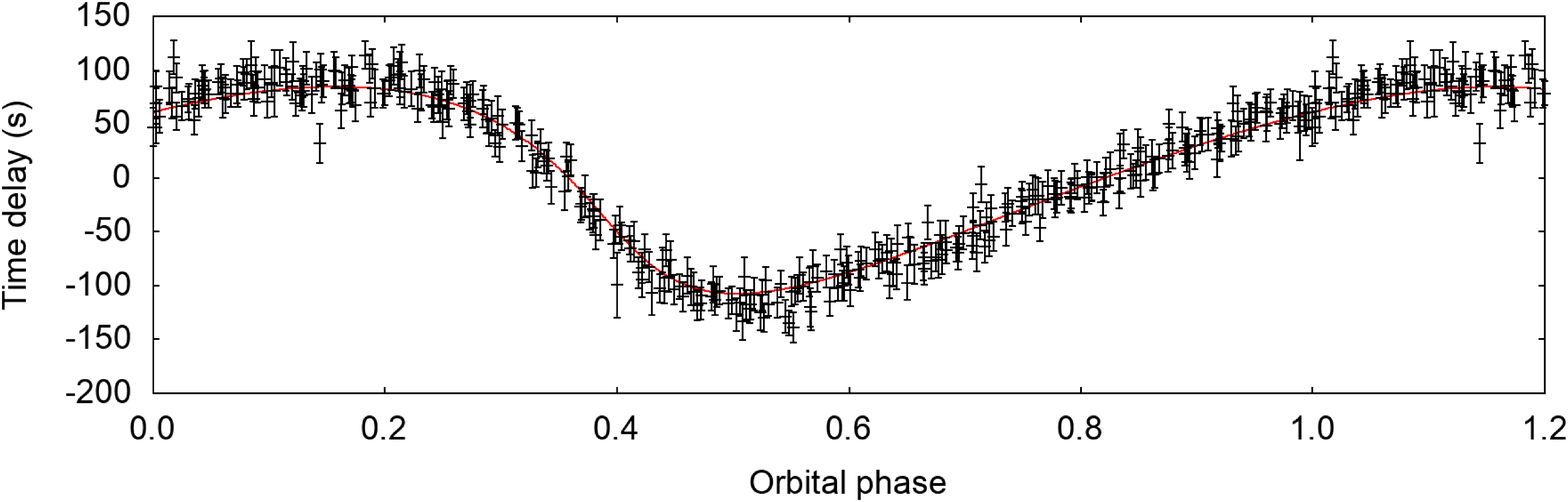}
\caption{The weighted average time delays folded on the orbital solution, shown with the best-fitting orbit computed using these time delays, only. The phase is calculated with respect to the time-stamp of the first time-delay measurement, BJD=2454955.53140. The $\chi^2/N$ of the fit was 0.75.}
\label{fig:TD_orbit}
\end{figure}

A tunable parameter in the PM method is the segment size over which the time delay is measured. Shorter segments offer finer sensitivity to the orbit at periastron, but lead to larger uncertainties per measurement and poorer frequency resolution in the Fourier transform \citep{murphyetal2014}. We experimented with segment sizes of 2, 4, 6, 8 and 10\,d and found 4\,d to be optimal for this target. We used the nine strongest Fourier peaks in the high-pass filtered light curve for the PM analysis, carefully avoiding Nyquist aliases \citep{murphyetal2013}, resulting in the time delays shown in Fig.\,\ref{fig:td_err_bar}. We ran the Markov-chain Monte-Carlo (MCMC) code of \citet{murphyetal2016b} on the weighted-average time delay to obtain the orbital parameters given in Table\:\ref{tab:orbit}. The orbital solution is shown in Fig.\,\ref{fig:TD_orbit}.

\subsection{Spectroscopic orbit determination}

Radial velocities (RVs) were determined by cross-correlating the spectra with a well-matching synthetic template spectrum from the extensive spectral library of \citet{mun05}. The calculated cross-correlation functions (CCFs) were fitted with two-component Gaussian functions, whose centroids gave the RVs for each component. Deviation from the Gaussian shape typically occurred around 100--150~km\,s$^{\rm -1}$ away from the maxima.

\begin{figure*}
\includegraphics[width=\linewidth]{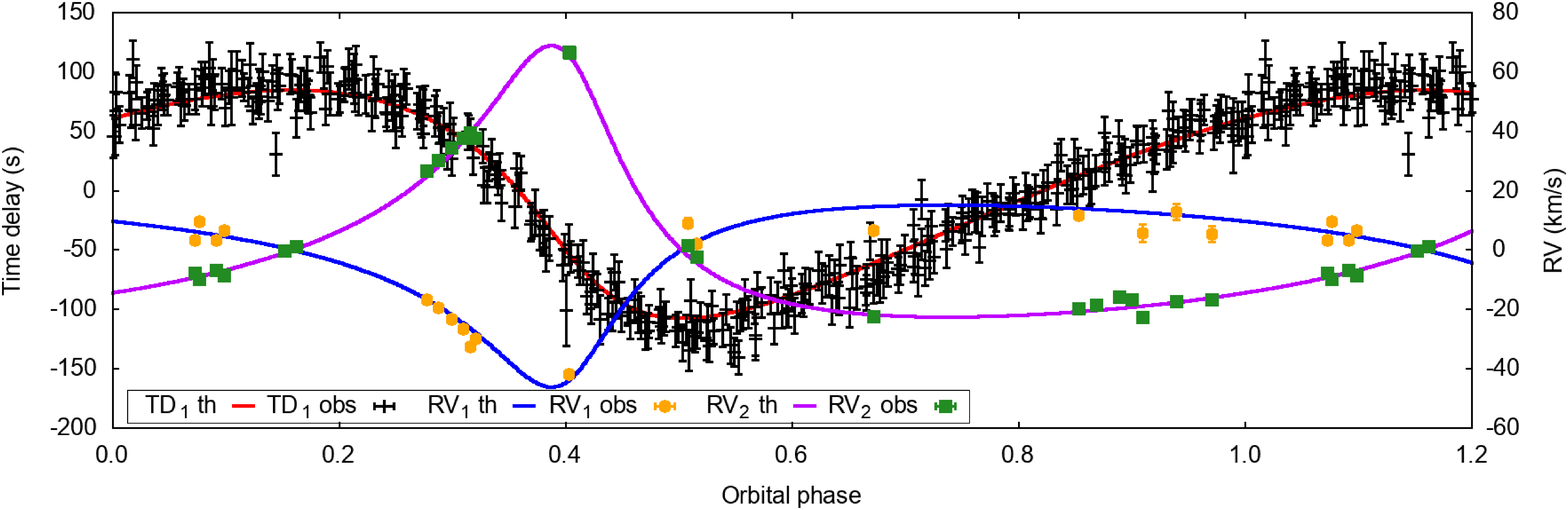}
\caption{Combined RV and time-delay data set for KIC~5709664, with the best fitting solution plotted as solid lines. The $\chi^2/N$ of this combined fit was 2.55.}
\label{fig:combined_orbit}
\end{figure*}

After extracting the RVs, we computed the RV curve. For this we used our own MCMC-based RV fitting code, implementing the Metropolis-Hastings algorithm. The RVs and the fit for each component are plotted in Fig.\:\ref{fig:combined_orbit} and the corresponding parameters are listed in Table\:\ref{tab:orbit}.

\subsection{A combined approach}
\label{ssec:combined}

The RV and PM methods are complementary. When used together, they offer a much longer observational span that allows the orbital parameters to be refined. They also allow the time delays to be allocated to one of the two components of the RV curve. While RVs constitute a near-instantaneous measurement of the orbit, our time-delay data used a 4-d integration and therefore undersample the orbit at periastron. We used the correction described by \citet{murphyetal2016b} to account for this.

We found that the  RV$_2$ velocities belong to the pulsating star for which the time delays were measured. To refine the orbit, we ran the PM MCMC algorithm on the joint data set without applying any additional weights, resulting in the parameters in the final column of Table\:\ref{tab:orbit}.

\begin{table*}
\centering
\caption{Orbital parameters for the KIC\,5709664 system. Only one component pulsates, so the time-delay (TD) orbit cannot extract parameters for the second star. The RV and TD+RV orbits used velocities from both components. The semi-major axes are given in light seconds, where 1 light second = 299\,792\,458\,m $\approx$ 0.431\,R$_{\odot}$ $\approx $ 0.002 au.}
\label{tab:orbit}
\begin{tabular}{c c r@{}l r@{}l r@{}l}
\toprule
\multicolumn{1}{c}{Parameter} & units & \multicolumn{2}{c}{TD only} & \multicolumn{2}{c}{RV only} & \multicolumn{2}{c}{TD + RV}\\
\midrule
\vspace{1.5mm}
$P_{\rm orb}$ & d & $95.024$&$^{+0.031}_{-0.034}$ & $94.9735$&$^{+0.023}_{-0.023}$ & $94.9854$&$^{+0.0058}_{-0.0057}$\\
\vspace{1.5mm}
$e$ & & $0.51$&$^{+0.020}_{-0.021}$ & $0.57$&$^{+0.015}_{-0.015}$ & $0.547$&$^{+0.0036}_{-0.0037}$\\
\vspace{1.5mm}
$\varpi$ & rad & $0.398$&$^{+0.037}_{-0.028}$ & $0.354$&$^{+0.034}_{-0.034}$ & $0.388$&$^{+0.013}_{-0.0096}$\\
\vspace{1.5mm}
$t_{\rm p}$ & d & $2\,454\,993.33$&$^{+0.57}_{-0.49}$ & $2\,454\,993.42$&$^{+0.66}_{-0.66}$ & $2\,454\,993.8$&$^{+0.21}_{-0.19}$\\
\vspace{1.5mm}
$f(m_1,m_2,\sin i)$ & M$_{\odot}$ & $0.1872$&$^{+0.0096}_{-0.0090}$ &$0.1566$&$^{+0.009}_{-0.009}$ & $0.1666$&$^{+0.0045}_{-0.0046}$\\
\vspace{1.5mm}
$f(m_2,m_1,\sin i)$ & M$_{\odot}$ & \multicolumn{2}{c}{---}  &$0.5438$&$^{+0.028}_{-0.028}$ & $0.5638$&$^{+0.0153}_{-0.0155}$\\
\vspace{1.5mm}
$a_1 \sin i / c$ & s & $116.3$&$^{+2.0}_{-1.9}$ & $109.5$&$^{+3.3}_{-3.3}$ & $111.9$&$^{+1.0}_{-1.0}$\\
\vspace{1.5mm}
$a_2 \sin i / c$ & s & \multicolumn{2}{c}{---} & $165.8$&$^{+3}_{-3}$ & $166.8$&$^{+0.65}_{-0.64}$\\
\vspace{1.5mm}
$q$ & & \multicolumn{2}{c}{---} & $0.6602$&$^{+0.065}_{-0.065}$ & $0.6706$&$^{+0.0065}_{-0.0068}$\\
\vspace{1.5mm}
$K_1$ & km\,s$^{-1}$ & $31.05$&$^{+0.94}_{-0.88}$ & $30.62$&$^{+1.02}_{-1.02}$ & $30.67$&$^{+0.31}_{-0.31}$\\
\vspace{1.5mm}
$K_2$ & km\,s$^{-1}$ & \multicolumn{2}{c}{---} &$46.37$&$^{+1.08}_{-1.08}$ & $45.74$&$^{+0.20}_{-0.20}$\\
\vspace{1.5mm}
$v_{\gamma}$ & km\,s$^{-1}$ & \multicolumn{2}{c}{---} & $-22.01$&$^{+0.3}_{-0.3}$ & $-21.94$ & \\
\bottomrule
\end{tabular}
\end{table*}

\section{Frequency analysis}
\label{sec:freqs}

\subsection{Extraction of p modes}

We used the orbit determined in Sect.\,\ref{ssec:combined} to correct the light arrival times to the barycentre of the binary system, so that there was no longer any frequency modulation of any of the pulsation modes. This leads to the cleanest Fourier spectrum, with no orbital sidelobes on any pulsation peaks, for analysis of the pulsation content.

\begin{figure}
\includegraphics[width=\columnwidth]{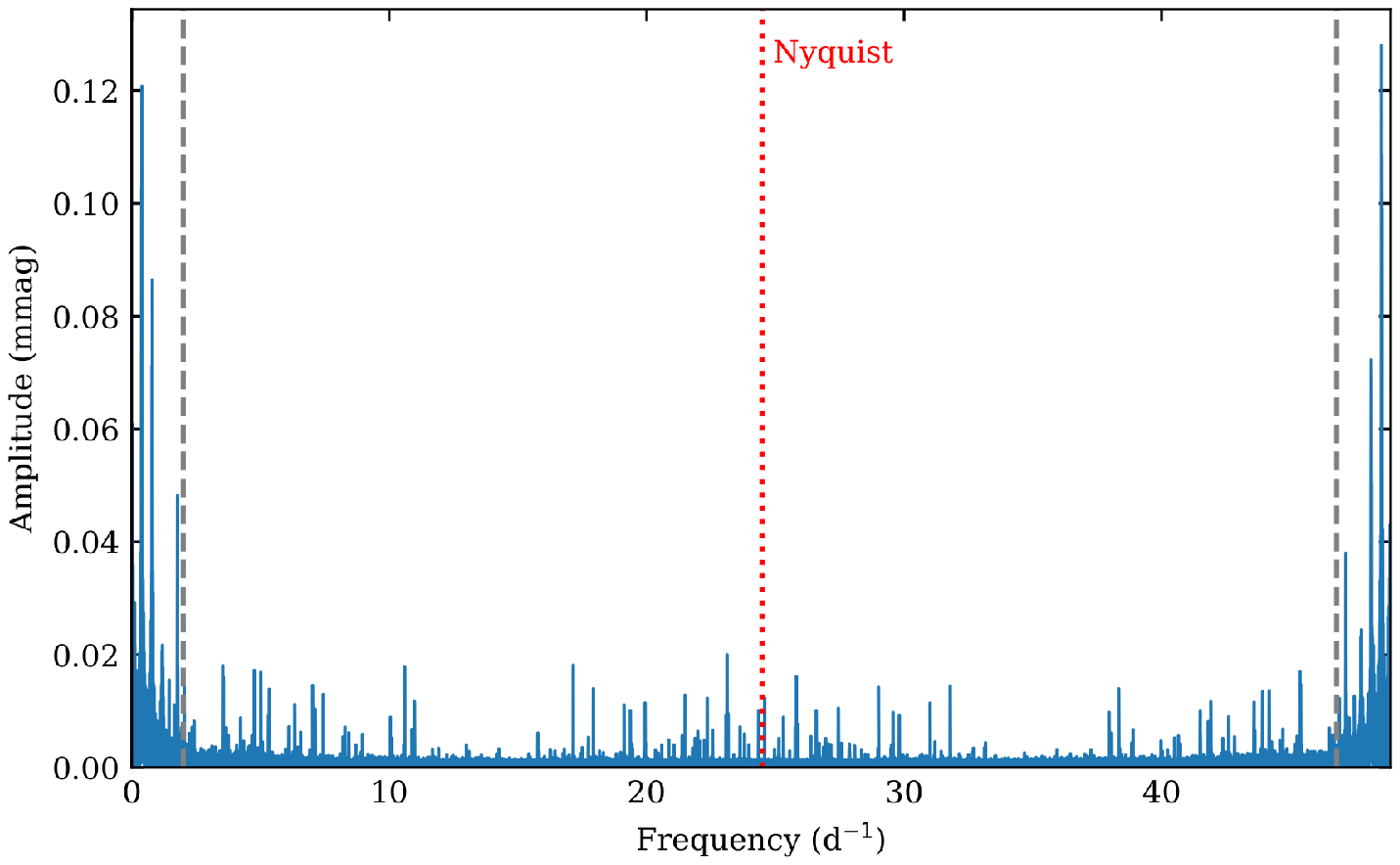}
\caption{The Fourier resdiuals after the 38-frequency fit. The dotted red line is the Nyquist frequency and the dashed grey lines delimit the frequency region considered for frequency extraction. The low-frequency region is shown in Fig.\,\ref{fig:low_freqs}.}
\label{fig:residuals}
\end{figure}

We used the {\small PERIOD04} software \citep{len2004} to identify pulsation frequencies in these data and to fit them to the light curve in order to compute their amplitudes and phases. We used the `super-Nyquist asteroseismology' method \citep{murphyetal2013} to distinguish the Nyquist aliases from the real oscillation frequencies without recourse to the \textit{Kepler} SC data. Below 2\,d$^{-1}$ there are some incoherent peaks in the Fourier transform, presumably from spots and rotation on one of the stars (discussed below), so we did not extract pulsation frequencies below 2\,d$^{-1}$. Frequencies were extracted in this manner down to 20\,$\mu$mag, resulting in the 38 frequencies, amplitudes and phases shown in Table\:\ref{tab:freqs}. While further significant peaks could be extracted below the 20\,$\mu$mag threshold, these are of decreasing significance and it is not possible to model so many frequencies in $\delta$\,Sct stars at present, so we ceased frequency extraction here. Our lowest-amplitude frequency has a 10$\sigma$ significance. The Fourier residuals after our frequency extraction are shown in Fig.\,\ref{fig:residuals}.

\begin{table}
\centering
\caption{The 38 extracted frequencies in descending amplitude order. The time zero-point for the phases is BJD=2\,455\,688.77. All amplitudes have the same uncertainty.}
\begin{tabular}{r@{\,$\pm$\,}l c r@{\,$\pm$\,}l}
\toprule
\multicolumn{2}{c}{Frequency} & Amplitude & \multicolumn{2}{c}{Phase} \\
\multicolumn{2}{c}{d$^{-1}$} & mmag $\pm$ 0.0024 & \multicolumn{2}{c}{($-\pi$ to $+\pi$)} \\ 
\midrule
$ 19.440055 $&$ 0.000001 $&$ 0.765 $&$ 2.093 $&$ 0.003 $\\
$ 16.259601 $&$ 0.000001 $&$ 0.712 $&$ -0.846 $&$ 0.003 $\\
$ 22.558025 $&$ 0.000002 $&$ 0.388 $&$ 2.446 $&$ 0.006 $\\
$ 19.123847 $&$ 0.000003 $&$ 0.294 $&$ -0.785 $&$ 0.008 $\\
$ 27.875417 $&$ 0.000003 $&$ 0.274 $&$ -2.343 $&$ 0.009 $\\
$ 22.075406 $&$ 0.000003 $&$ 0.265 $&$ 0.075 $&$ 0.009 $\\
$ 17.919903 $&$ 0.000005 $&$ 0.194 $&$ -1.515 $&$ 0.013 $\\
$ 18.452354 $&$ 0.000005 $&$ 0.187 $&$ -0.211 $&$ 0.013 $\\
$ 25.555143 $&$ 0.000005 $&$ 0.186 $&$ 0.918 $&$ 0.013 $\\
$ 20.883914 $&$ 0.000006 $&$ 0.151 $&$ -0.299 $&$ 0.016 $\\
$ 13.902632 $&$ 0.000006 $&$ 0.144 $&$ -1.392 $&$ 0.017 $\\
$ 24.424301 $&$ 0.000007 $&$ 0.141 $&$ -1.782 $&$ 0.017 $\\
$ 16.311322 $&$ 0.000008 $&$ 0.111 $&$ 2.630 $&$ 0.022 $\\
$ 2.752339 $&$ 0.000009 $&$ 0.107 $&$ 0.511 $&$ 0.023 $\\
$ 20.015493 $&$ 0.000009 $&$ 0.100 $&$ -2.862 $&$ 0.024 $\\
$ 24.557842 $&$ 0.000010 $&$ 0.091 $&$ 2.590 $&$ 0.027 $\\
$ 2.139325 $&$ 0.000013 $&$ 0.072 $&$ -1.483 $&$ 0.034 $\\
$ 15.663759 $&$ 0.000014 $&$ 0.067 $&$ -0.801 $&$ 0.036 $\\
$ 12.236958 $&$ 0.000015 $&$ 0.062 $&$ -2.568 $&$ 0.039 $\\
$ 15.636912 $&$ 0.000015 $&$ 0.061 $&$ -2.276 $&$ 0.040 $\\
$ 3.679324 $&$ 0.000018 $&$ 0.051 $&$ 2.898 $&$ 0.047 $\\
$ 22.237565 $&$ 0.000018 $&$ 0.051 $&$ 1.360 $&$ 0.047 $\\
$ 22.252450 $&$ 0.000019 $&$ 0.048 $&$ 0.653 $&$ 0.050 $\\
$ 19.907823 $&$ 0.000019 $&$ 0.047 $&$ -2.338 $&$ 0.051 $\\
$ 3.361601 $&$ 0.000021 $&$ 0.043 $&$ -2.718 $&$ 0.057 $\\
$ 5.345200 $&$ 0.000024 $&$ 0.038 $&$ -1.275 $&$ 0.063 $\\
$ 24.374664 $&$ 0.000024 $&$ 0.038 $&$ 1.902 $&$ 0.064 $\\
$ 23.471627 $&$ 0.000026 $&$ 0.036 $&$ -2.328 $&$ 0.068 $\\
$ 12.296338 $&$ 0.000027 $&$ 0.034 $&$ -1.905 $&$ 0.071 $\\
$ 23.529187 $&$ 0.000029 $&$ 0.032 $&$ 2.360 $&$ 0.076 $\\
$ 3.563488 $&$ 0.000029 $&$ 0.032 $&$ 1.171 $&$ 0.077 $\\
$ 21.697655 $&$ 0.000029 $&$ 0.031 $&$ -2.938 $&$ 0.078 $\\
$ 2.865545 $&$ 0.000031 $&$ 0.030 $&$ 0.025 $&$ 0.082 $\\
$ 24.917230 $&$ 0.000032 $&$ 0.029 $&$ 1.558 $&$ 0.084 $\\
$ 4.955035 $&$ 0.000033 $&$ 0.028 $&$ -1.029 $&$ 0.088 $\\
$ 4.195776 $&$ 0.000036 $&$ 0.025 $&$ -2.610 $&$ 0.096 $\\
$ 27.160243 $&$ 0.000038 $&$ 0.024 $&$ -0.892 $&$ 0.100 $\\
$ 27.867651 $&$ 0.000038 $&$ 0.024 $&$ 3.101 $&$ 0.100 $\\
\bottomrule
\end{tabular}
\label{tab:freqs}
\end{table}

We note that the strongest two frequencies, at 19.4401 and 16.2596\,d$^{-1}$, have a period ratio of 0.836, consistent with the expected ratio for the third and second radial overtone modes \citep{smolecetal2017}.

\subsection{Low frequencies}
\label{ssec:low_freqs}

\begin{figure}
\includegraphics[width=\columnwidth]{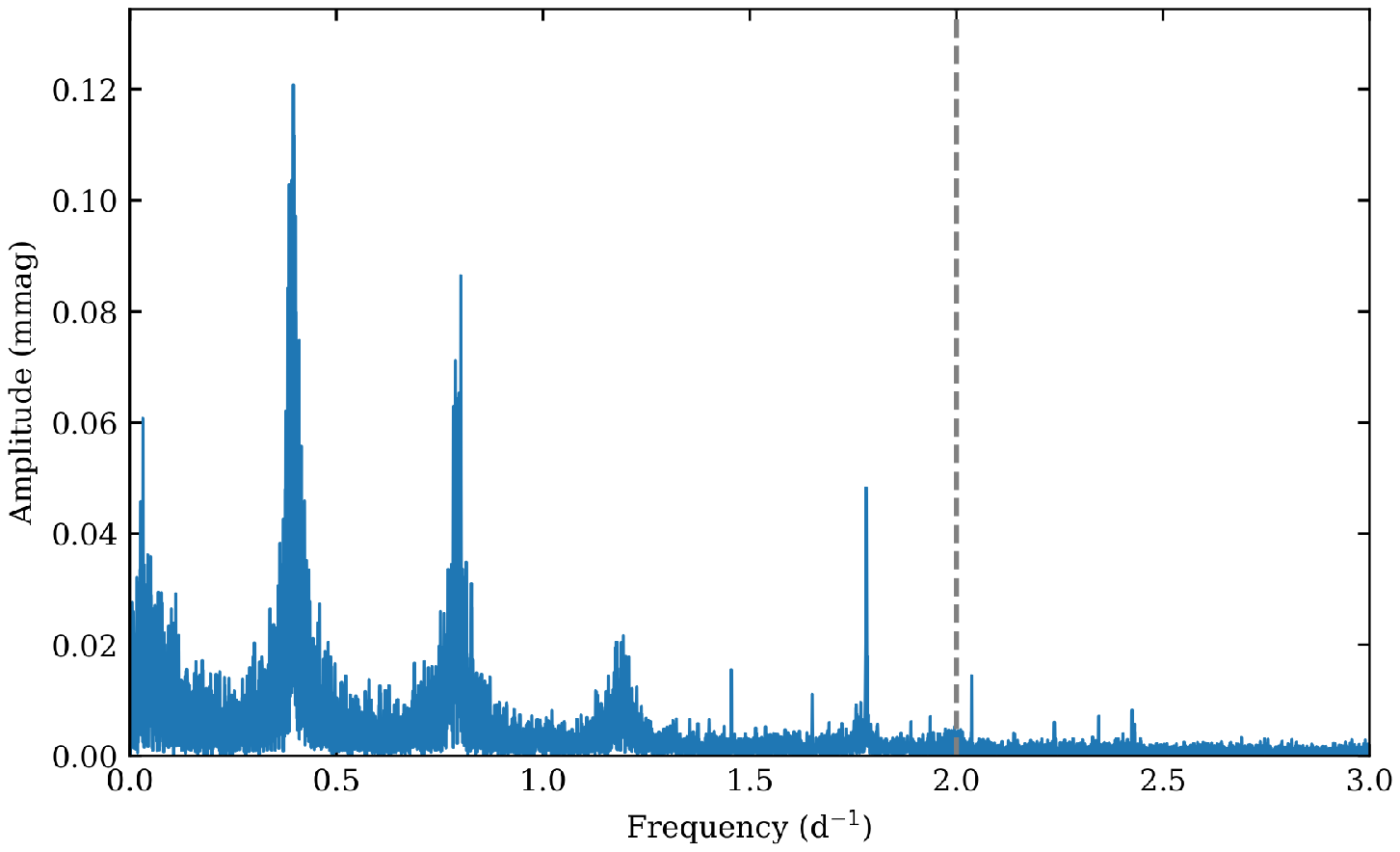}
\caption{The Fourier transform of the low-frequency region, after the 38-frequency fit. The dashed grey line delimit the frequency region above which peaks were extracted. The three broad humps might be from spots and rotation on the secondary component. The feature at 1.78\,d$^{-1}$ could be attributed to r\:modes.}
\label{fig:low_freqs}
\end{figure}

\begin{figure}
\includegraphics[width=\columnwidth]{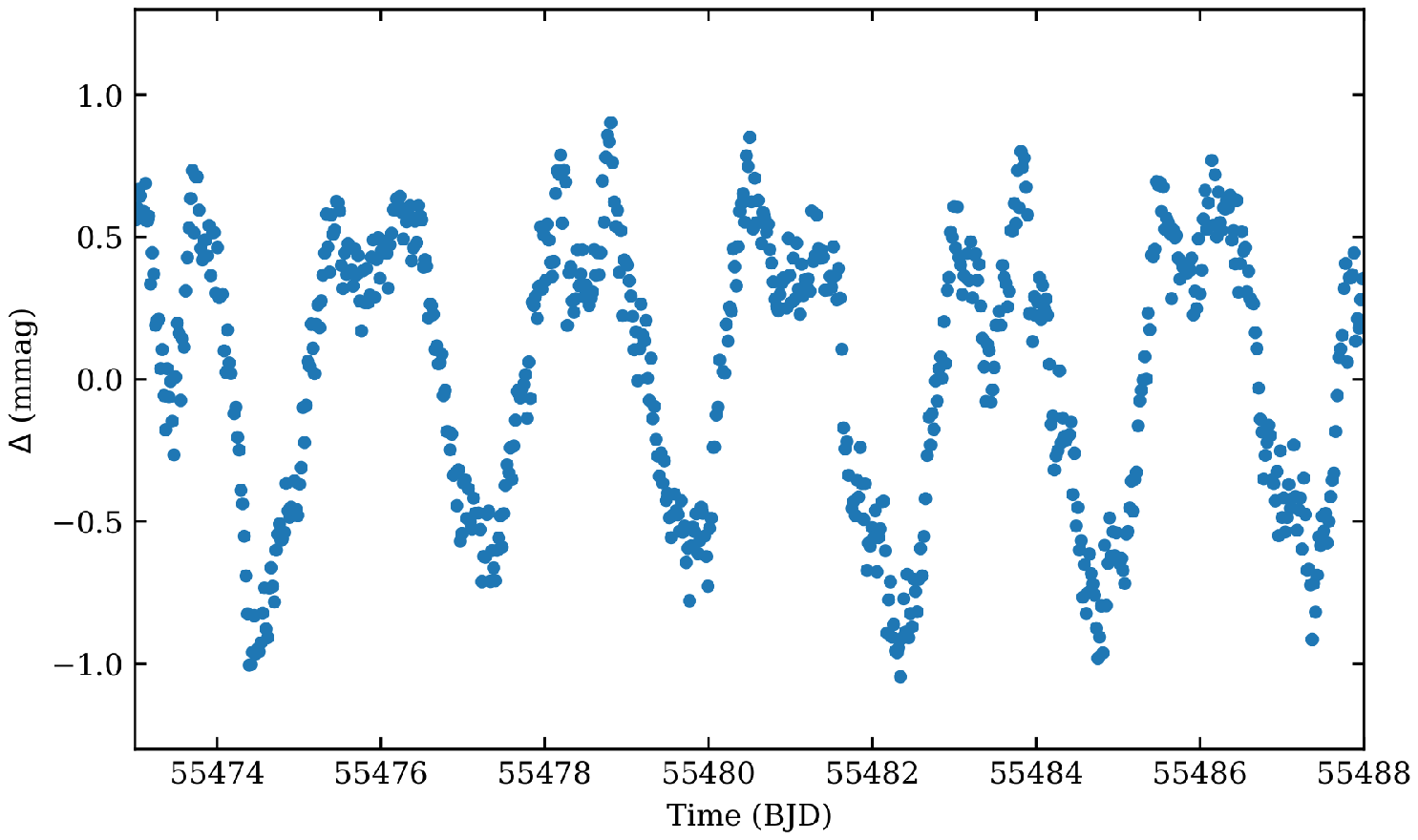}
\caption{A 15-d segment of the residual light curve (with the 38-frequency fit subtracted), showing the low-frequency variability that we attribute to starspots on the cooler companion.}
\label{fig:spots}
\end{figure}

At low frequencies there are broad power excesses (Fig.\,\ref{fig:low_freqs}), which were not extracted with the p\:modes. They might be attributed to spots on a rotating star. Since the primary is a $\delta$\,Sct star with a mass likely in the region 1.5--2.0\,M$_{\odot}$, the measured mass ratio of 0.67 (Table\:\ref{tab:orbit}) gives a companion mass in the range 1.0--1.34\,M$_{\odot}$, which should have a surface convection zone capable of generating sun-like starspots. As these starspots migrate over the surface and as the star rotates, these cause brightness variations. Such starspot-induced variability has been detected in thousands of \textit{Kepler} light curves \citep{nielsenetal2013,mcquillanetal2014}. We show a 15-d segment of the residual light curve that illustrates the typical rotational modulation from sun-like starspots in Fig.\,\ref{fig:spots}.

Also evident in the low-frequency region of the Fourier transform (Fig.\,\ref{fig:low_freqs}) at 1.78\,d$^{-1}$ is a signature of r\:modes, which appear to be ubiquitous among A stars \citep{saioetal2018}. The morphology is a `hump and spike', of which the spike is believed to be the rotation frequency of the star, mechanically generating r\:modes that are visible as the hump. If the r\:modes are attributable to the primary, the spike suggests a rotation frequency of 1.782\,d$^{-1}$ and a corresponding period of 0.561\,d for this component, consistent with typical rotation periods of A-type stars \citep{royeretal2007}. Similarly,  if we take the first Fourier hump of the secondary's starspot signature as the rotation frequency, noting that there are potential caveats to this interpretation \citep{mcquillanetal2014,angusetal2018}, then we obtain a rotation frequency for the secondary of 0.395\,d$^{-1}$, corresponding to a period of 2.53\,d. While this is relatively rapid for a star of this mass, the shorter pre-main-sequence phase of the more massive primary will have limited early rotational braking via disk-locking, and left the secondary rotating relatively rapidly.

\section{Conclusions and summary} \label{sec:sum}

 We analysed {\it Kepler} long-cadence data of the hybrid $\delta$\,Sct/$\gamma$\,Dor star KIC~5709664. The frequency analysis of the photometric data revealed Fourier peaks with sidelobes, suggesting binary orbital motion.  We used the nine strongest Fourier peaks for the Phase Modulation (PM) analysis and determined the orbital parameters.

We obtained 25 medium- and high-resolution spectra at 6 observatories, which unambiguously confirmed the binary nature. The calculated cross-correlation functions were fitted with a two-component Gaussian function and the RV curves of each binary component were fitted to measure the orbital parameters. 

Although the parameters determined by the two independent methods are in good agreement, we performed a combined radial velocity and phase modulation approach to determine more accurate orbital parameters that are listed in the final column of Table\,\ref{tab:orbit}. We found that the pulsating star for which the time delays are measured is the component to which the RV$_{2}$ velocities belong.

We performed frequency analysis of the data, and extracted 38 frequencies with amplitudes greater than 20\,$\mu$mag. At low frequencies, we found broad power excesses which are likely attributed to spots on the rotating surface of the primary component. Since the primary is a $\delta$\,Sct star with a mass likely in the region 1.5--2.0\,M$_{\odot}$, the measured mass ratio of 0.67 gives a companion mass in the range 1.0--1.34\,M$_{\odot}$, which should have a surface convection zone capable of generating sun-like starspots. We also found a signature of r\:modes, presumably belonging to the primary. We inferred rotation periods of 0.56 and 2.53\,d for the primary and secondary, respectively.

\section*{Acknowledgements}

This project has been supported by the Hungarian NKFI Grants K-113117, K-115709,  K-119517 and KH-130372 of the Hungarian National Research, Development and Innovation Office. AD was supported by the \'UNKP-18-4 New National Excellence Program of the Ministry of Human Capacities and the J\'anos Bolyai Research Scholarship 
of the Hungarian Academy of Sciences.. AD would like to thank the City of Szombathely for support under Agreement No. 67.177-21/2016. SJM is a DECRA fellow supported by the Australian Research Council (grant number DE180101104). This project has been supported by the Lend\"ulet Program of the Hungarian Academy of Sciences, project No. LP2018-7/2018. SzM has been supported by the Premium Postdoctoral Research Program of the Hungarian Academy of Sciences. HL was supported by the DFG grant LE 1102/3-1.

Based on observations obtained with the APO 3.5-m telescope, which is owned and operated by the Astrophysical Research Consortium. IRAF is distributed by the National Optical Astronomy Observatories, which are operated by the Association of Universities for Research in Astronomy, Inc., under cooperative agreement with the National Science
Foundation.







\bsp	
\label{lastpage}
\end{document}